\documentclass[twocolumn,prl,showpacs]{revtex4}
\usepackage{epsfig}
\begin{document}
\author{M. W. Mitchell$^{1}$, C. W. Ellenor$^{1}$ , S. 
Schneider$^{2}$ and A. M. Steinberg$^{1}$}
\affiliation{$^{1}$Department of Physics, University of Toronto, 60 
St.~George St., Toronto, ON M5S 1A7, Canada \\
$^{2}$Chemical Physics Theory Group, Department of Chemistry, 
University of Toronto, 80 St. George St., Toronto, ON 
M5S 3H6, Canada}
\newcommand{\DateWritten}{\today}
\title{
Diagnosis, prescription and 
prognosis of a Bell-state 
filter by quantum process tomography.}
\newcommand{\FigWidth}{3.0in}
\begin{abstract} 
Using a Hong-Ou-Mandel
interferometer, we apply the techniques of quantum process tomography
to characterize errors and decoherence in a prototypical two-photon
operation, a singlet-state filter.  The quantum process tomography
results indicate a large asymmetry in the process and also the required 
operation to correct for this asymmetry.  Finally, we quantify errors
and decoherence of the filtering operation after this modification.

%
\end{abstract}



\maketitle

\newcommand{\be}{\begin{equation}}
\newcommand{\ee}{\end{equation}}
\newcommand{\bea}{\begin{eqnarray}}
\newcommand{\eea}{\end{eqnarray}}
\newcommand{\ket}[1]{\left|#1\right>}
\newcommand{\bra}[1]{\left<#1\right|}
\newcommand{\bk}{{\bf k}}
\newcommand{\etal}{{\em et al.}} 
\newcommand{\degree}{$^{\circ}$~}
\newcommand{\bigepsilon}{{\cal E}}
\newcommand{\polket}{}
{
Quantum computation promises exponential speedup in the solution of
difficult problems such as factoring large numbers and simulating
quantum systems \cite{DiVincenzo1995, EckertJozsa1996}.  In a quantum
computer single- and multiple-qubit operations drive
the system through a sequence of highly entangled states before the
result is finally measured.  A quantum computation is vulnerable to
errors and to environmental decoherence, which destroys the
entanglement.  Characterization of quantum operations including errors
and decoherence is a pressing issue for quantum information
processing \cite{Roadmap}, and is possible by the technique of {\em
quantum process tomography} (QPT) \cite{Poyatos1997,ChuangNielsen1997}. 
QPT has been demonstrated for single
qubits \cite{DeMartini2002,Altepeter2003} and for mixed ensembles of
two-qubit systems \cite{Braunstein1999} in NMR \cite{Childs2001}.   Here
we present QPT of an entanglement-generating two-qubit operation, the
partitioning of photons by a beamsplitter in a Hong-Ou-Mandel (HOM)
interferometer.  Our characterization reveals large imperfections in
the process and indicates the appropriate remedy.  Finally, we extend
the QPT results to predict the accuracy of the process, once repairs
are carried out.

}

Multi-qubit operations on photons, once thought to require very large
optical nonlinearities, can now be performed with linear optical
elements such as wave-plates and beamsplitters coupled with the highly
nonlinear process of photodetection.  This idea is exploited in
schemes for linear optics quantum computation
\cite{KLM2001,Franson2002,Gilchrist2003} and to generate
multi-photon entangled states \cite{KokDowling2002, Fiurasek2002}.  The
schemes are probabilistic and employ {\em post-selection}: the
photodetection signals indicate when the correct operation has taken
place.  The HOM effect plays a central role in these proposals, and
itself is a prototypical example of a post-selected multi-qubit
operation; it generates correlations and entanglement without optical
nonlinearities.  The HOM effect has been used to produce entangled
states for Bell inequality tests \cite{Ou1988,Shih1988} and to make
probabilistic Bell state measurements for quantum teleportation and
entanglement swapping \cite{Bouwmeester1997,Pan1998}.

In the HOM effect, two photons meeting at a 50/50 beamsplitter can
leave by different output ports only if they are in some way
distinguishable \cite{HongOuMandel}.  We use photon
pairs which are indistinguishable in wavelength, spatial mode and
arrival time at the beamsplitter, leaving only the polarization to
(possibly) distinguish them.  By detecting photons leaving from
different output ports, we post-select an entangled polarization
state.  Ideally, the process acts as a filter for the Bell singlet
state $\polket{\Psi^{-}} = (\polket{HV} - \polket{VH})/\sqrt{2}$, in 
which the photons have orthogonal polarizations in any basis.  
In any real apparatus this process will include errors and 
decoherence.  Using the 
techniques of QPT, we determine how the polarization state, more
specifically the $4\times 4$ density matrix $\rho$ which describes an
arbitrary two-photon mixed state, changes in passing the beamsplitter.  
In general, $\rho$ will change as
$\rho^{(in)} \rightarrow \rho^{(out)} = \bigepsilon(\rho^{(in)})$,
where $\bigepsilon$ is the {\em superoperator}, a linear mapping from 
input density matrices to output density matrices.  The superoperator
completely characterizes the effect on the system, including 
coherent evolutions, decohering interactions with the environment,
and loss.  

\begin{figure}[h]
\centerline{\epsfig{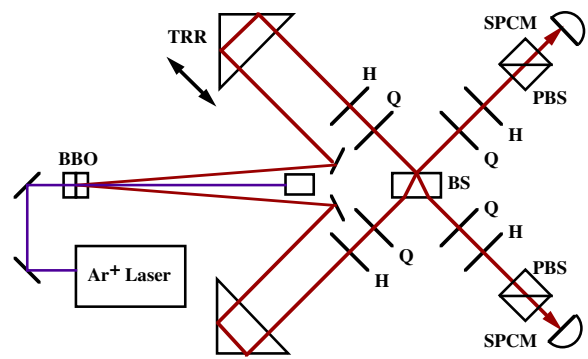}}
\caption{Schematic of experimental setup.  BBO, $\beta$-barium borate 
crystals, H half-wave plate, Q quarter wave plate, BS beamsplitter, 
PBS polarizing beamsplitter, SPCM single-photon counting module, TRR 
translatable retro-reflector. }
\end{figure}


\newcommand{\rv}{\vec{\rho}}
\newcommand{\sm}{{\bf M}}

We use a HOM
interferometer constructed to produce arbitrary input
polarizations and detect arbitrary output polarizations.  The
experimental setup is shown schematically in Fig. 1.  A 7 mW beam of
351.1 nm light from an argon-ion laser 
illuminates a pair of 0.6 mm thick $\beta$-barium borate 
crystals, cut for degenerate
downconversion at a half-opening angle of 3.3\degree.  Pairs of
downconversion photons at 702.2 nm emerge from the crystals 
vertically polarized.
This initial polarization state can be rotated into any input product state
by the state preparation 
half- and quarter-wave plates immediately before the central
beamsplitter.  The downconversion beams meet the beamsplitter at 45\degree 
incidence.  The beamsplitter itself \cite{CVIBeamsplitter}
consists of a multi-layer dielectric coating on a glass substrate,
with an anti-reflection coated back face.
  
Polarization analyzers
consisting of a quarter- and a half-waveplate before a polarizing
beamsplitter are used to select an arbitrary product state.
Photons which pass the analyzers
are detected by single-photon counting modules and individual and
coincidence detection rates are registered on a computer.
Downconversion beams were aligned to overlap both spatially and
temporally on the beamsplitter, giving a HOM dip visibility of 90
$\pm$ 5\% for both horizontal and vertical input polarizations.  The
process tomography measurements described below were performed at the
center of this dip.



We prepare 16 linearly independent input states $\{
\rho^{(in)}_{i} \} $ and measure the corresponding outputs $\{
\rho^{(out)}_{i} \}$.  
%
%
The inputs \cite{JKMW} are the pure states $\rho_{i}^{(in)} = 
\ket{\psi_{i}}\bra{\psi_{i}}$ where
\bea
\{ \polket{\psi_{1}},\ldots,\polket{\psi_{16}} \} &=& 
\{
\polket{HH},
\polket{HV},
\polket{VV},
\polket{VH},
\nonumber \\
& &
\polket{RH},
\polket{RV} 
\polket{DV},
\polket{DH},
\nonumber \\
& &
\polket{DR},
\polket{DD},
\polket{RD},
\polket{HD},
\nonumber \\
& &
\polket{VD},
\polket{VL},
\polket{HL},
\polket{RL} 
\}
\eea
and the polarizations are horizontal $H$, vertical $V$, 
diagonal $D = (H+V)/\sqrt{2}$, 
right circular  $R=(H-iV)/\sqrt{2}$ 
and left circular  $L=(H+iV)/\sqrt{2}$.  
A single output $\rho_{i}^{(out)}$ can be found by making 
projective measurements onto the sixteen states $\{\psi_{j}\}$
.  
The coincidence rates for these measurements are $R_{ij} = R_{0}
{\mathrm{Tr}}[\rho_{i}^{(out)}\ket{\psi_{j}}\bra{\psi_{j}}]$, where
$R_{0}$ is the constant rate of downconversion at the crystals.
Note that we use non-normalized output density matrices,
i.e. ${\mathrm{Tr}}[\rho^{(out)}] \le 1$, because photon pairs can be
lost in the process.  Absorption and scattering losses are small, but
post-selection necessarily removes a significant fraction of the pairs
for most input states.


\begin{figure}[h]
\centerline{\epsfig{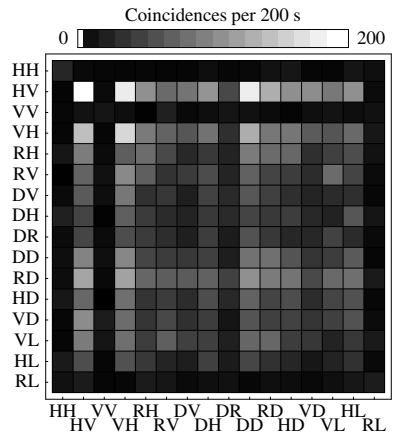}}
\caption{Coincidence rates.  Brightness
indicates the count rate observed in a given two-photon polarisation
state (horizontal axis) for a given input polarisation state 
(vertical axis).}
\end{figure}

\begin{figure}[h]
\centerline{\epsfig{width=\FigWidth,figure=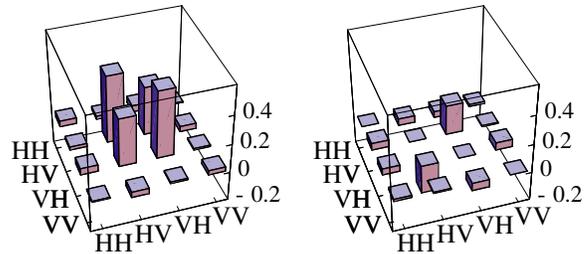}}
\caption{Output density matrix (normalized) for an input state 
of $HV$.
Left graph shows $\mathrm{Re}[\rho^{(out)}]$, right graph shows
$\mathrm{Im}[\rho^{(out)}]$.
}
\end{figure}

The measured coincidence rates $R_{ij}$ are shown in Fig.
2.  As expected for a filter, the output has similar
polarization characteristics for all inputs, but not all are 
equally transmitted, e.g., $\polket{HH}$ and $\polket{VV}$ are blocked. 
A typical output density matrix, 
reconstructed using maximum-likelihood estimation \cite{JKMW} 
is shown in Fig. 3.  The large coherence 
between $\polket{HV}$ and $\polket{VH}$ indicates that this is an entangled state,
with a concurrence \cite{Wooters1998, Coffman2000, JKMW} of $C = 0.89$.
The HOM effect is acting as an entangled-state filter, but the 
selected state is clearly not $\polket{\Psi^{-}}$, which has a real
density matrix and {\em negative} off-diagonal elements.   

\begin{figure}[h]
\centerline{\epsfig{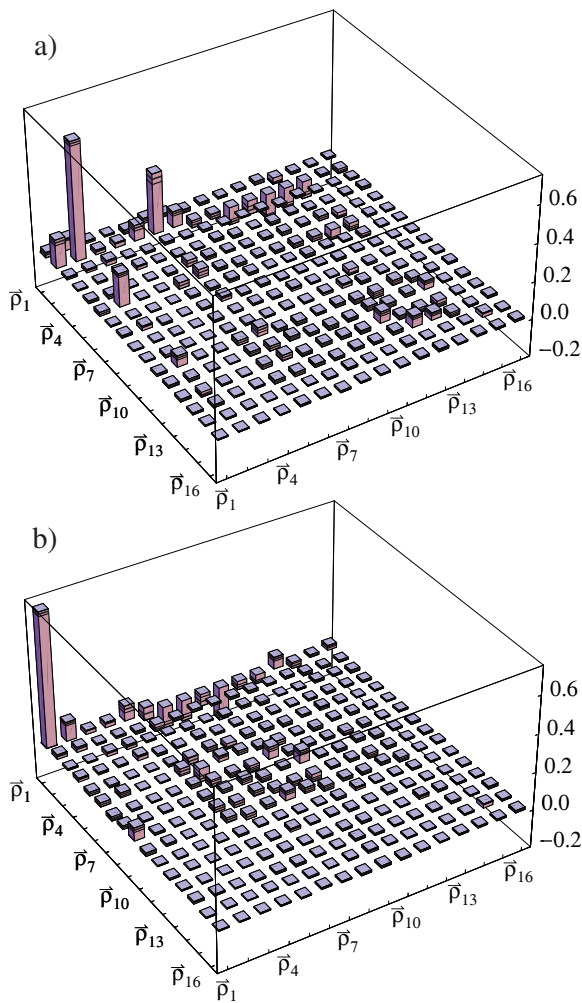}}
\caption{Reconstructed superoperators for the post-selected HOM 
process.  a) superoperator as measured, b) predicted superoperator 
after repair.  The matrix $\sm$ is shown, input density matrix 
elements at bottom, 
output  elements at left, where the density matrices are represented in
vector form (see text).  Horizontal stripes on the vertical bars indicate 
the the best estimated value and the statistical uncertainties.
  }
\end{figure}

We can understand this behaviour through the superoperator $\bigepsilon$.
For clarity, we work in the Bell state basis
$\{\polket{\Psi^{-}},\polket{\Psi^{+}},\polket{\Phi^{-}},\polket{\Phi^{+}}\}$
where $\polket{\Psi^{\pm}} = (\polket{HV} \pm \polket{VH})/\sqrt{2}$ and
$\polket{\Phi^{\pm}} =  (\polket{HH} \pm \polket{VV})/\sqrt{2}$.  We 
use a matrix representation for {$\cal{E}$}:  The
density matrix is written as a real 16-dimensional vector $\rv$ 
made from the independent coefficients of the non-normalized density 
matrix, i.e.,
$\rv =
(\rho_{11}, \ldots, \rho_{44}, \mathrm{Re}[\rho_{12}],
\mathrm{Im}[\rho_{12}], \mathrm{Re}[\rho_{13}], \ldots, 
\mathrm{Im}[\rho_{34}])^{T}$.  The superoperator is represented 
by a 
matrix
$\sm$ which acts as
\be
\rv^{(out)} = \sm \rv^{(in)}.
\label{eq:SuperMatrixDefinition}
\ee
In principle, $\sm$ could be found from this equation by a simple 
inversion, since we measured $R_{ij}$ for a basis set $\{\rho^{(in)}_{i}\}$.  
This procedure is 
sensitive to small errors and can produce a non-physical $\sm$, 
i.e., one which predicts non-physical (mathematically, 
non-positive semidefinite) $\rho^{(out)}$.  
Instead, we reconstruct $\sm$ by maximum-likelihood estimation within the space of
completely positive superoperators, i.e.~operators that map physical density
matrices to physical density matrices (see e.g. Sudarshan \cite{Sudarshan1961} for
the mathematical conditions on the mapping between density matrices). 
The reconstructed $\sm$ is shown in Fig. 4 a), with error estimates from
an ensemble of simulated datasets Poisson distributed around the measured 
data.  This matrix is
``normalized'' to give ${\mathrm{Tr}}[\rho_{out}] = 1/4$ when the input is a 
completely mixed state.  

We verify the accuracy of the reconstructed superoperator using the 
the input states $\polket{LL}$ and $\polket{RR}$, which were not used in 
the reconstruction process.  These states are used 
as input, both to equation (\ref{eq:SuperMatrixDefinition}) and
in the HOM interferometer. In both cases, prediction and the experiment 
result (again by maximum likelihood reconstruction) agree with fidelity of 
$97$\%.


The superoperator $\sm$ bears little resemblance to an ideal 
singlet-state filter, for which $M_{ij} = \delta_{i,1}\delta_{j,1}$.  Clearly 
the process is not performing the intended filtering operation.  
In fact, it is nearly a projection onto a different maximally 
entangled state \cite{Asymmetry}.  Written as a canonical 
Kraus operator 
sum \cite{Choi1975,Kraus1983}, the superoperator  
allows us to find this state directly.  In the sum
${\cal E}(\rho) = \sum_{l} 
\hat{K}_{l} \rho \hat{K}_{l}^{\dagger}$,
the leading 
operator $\hat{K}_{1}$ is very nearly a projector onto the state 
$\polket{\Psi^{-}_{\phi}} \equiv (\polket{HV} - \exp[i \phi] 
\polket{VH})/\sqrt{2}$ with $\phi = 0.84~\pi$. 
This immediately suggests a way to ``fix'' the non-ideal 
beamsplitter.  Adding a birefringent phase shifter which takes
$VH\rightarrow \exp[-i \phi]VH$ before the beamsplitter and the 
reverse operation afterward would give (nearly) a single-state filter.  

Finally, we can predict the behaviour of this ``fixed'' operation. 
The corresponding matrix $\sm$ is shown in Fig.  4b).  The large (1,1)
element indicates the filtering operation and the smaller nonzero
elements contribute to decoherence and other errors.  These errors
presumably arise from imperfect overlap of the downconverted beams and
residual imperfections in the beamsplitter.  They do not appear to
have a simple form, but we can gain some insight from some simple
measures, calculated using the superoperator.  
An unpolarized input (a completely mixed state) gives rise to an output 
that is 84\% $\polket{\Psi^{-}}$, or
an average polarization ratio (intensity of $\polket{\Psi^{-}}$  
versus average intensity of the other three Bell states) of
16:1.  This same output is entangled, with a concurrence
of $C = 0.70$, sufficient for a Bell inequality violation. 
Quantifying purity with the linear entropy $S_{L}$, which ranges from
0 for a pure state to 1 for a completely mixed state, the process
purifies the mixed state from $S_{L}=1$ to $S_{L}=0.37$.  We can also
ask how well the repaired filter maintains the coherence of an input. 
The pure input $\Psi^{-}$ is passed 75\% of the time and emerges
largely pure, with $S_{L} = 0.13$.  The state $\Psi^{+}$ is 13\%
passed with $S_{L} = 0.51$ and $\Phi^{\pm}$ are on average 6\% passed
with low purity $S_{L} = 0.88$.  Of course, different applications for
a singlet-state filter will have different requirements and different
figures of merit.  The superoperator we have found using QPT is more
general, a complete characterization suitable for evaluating any
proposed use.  It is also, we have seen, a useful diagnostic and
predictive tool.

We thank Daniel Lidar, Jeff Lundeen and Kevin Resch for assistance and
helpful discussions.   This
work was supported by the National Science and Engineering Research
Council of Canada, Photonics Research Ontario, the Canadian Institute
for Photonic Innovations and the DARPA-QuIST
program (managed by AFOSR under agreement No. F49620-01-1-0468).

\bibliographystyle{revtex}

\end{document}